# The periodic domino problem is undecidable in the hyperbolic plane

Maurice Margenstern[1]

Laboratoire d'Informatique Théorique et Appliquée, EA 3097,
Université de Metz, I.U.T. de Metz,
Département d'Informatique,
Île du Saulcy,
57045 Metz Cedex, France,
margens@univ-metz.fr

**Abstract.** In this paper, we consider the periodic tiling problem which was proved undecidable in the Euclidean plane by Yu. Gurevich and I. Koriakov, see [3]. Here, we prove that the same problem for the hyperbolic plane is also undecidable.

## 1 Introduction

A lot of problems deal with tilings. Most of them are considered in the setting of the Euclidean plane. A certain number of these problems turn out to be undecidable in this frame, thanks to the facility to simulate the computation of a Turing machine in this setting. The most famous case of such a problem is the **general tiling problem** proved to be undecidable by Berger in 1966, see [1]. In 1971, R. Robinson gave a simplified proof of the same result, see [13]. Sometimes, the general problem is simply called the **tiling problem**. The reason of these different names lies in the fact that several conditions were put on the problem, leading to different settings, and a dedicated proof was required each time when the problem turned out to be undecidable. Among these variations, the most well-known is the **origin-constrained** problem, proved to be undecidable by Wang in 1958, see [15].

The general tiling prolem consists in the following. Given a finite set of tiles $T$, is there an algorithm which says whether it is possible to tile the plane with copies of the tiles of $T$ or not? The **origin-constrained** problem consists in the same question to which a condition is appended: given a finite set of tiles $T$ and a tile $T_0 \in T$, is there an algorithm which says whether i is possible or not to tile the plane with copies of the tiles of $T$ or not, the first tile being $T_0$? In the general problem there is no condition on the first tile: it can be a copy of any tile of $T$.

There are a lot of variants of these problems and the reader is referred to [13], where an account is given on several such conditions.

The **periodic tiling problem** is a bit different question. Given a finite set of tiles $T$, is there a way to tile the plane with copies of $T$ in a **periodic** way? The problem was proved undecidable for the Euclidean plane by Yu. Gurevich and I. Koriakov in 1972. Now, it turns out that the notion of period is well defined in the Euclidean plane, but it is not clear how to define it in the hyperbolic plane. As many authors do, we shall consider that a tiling of the hyperbolic plane is periodic if it is unchanged under a non trivial shift.

The general tiling problem for the hyperbolic plane was raised by R. Robinson in his 1971 paper, see [13]. In 1978, R. Robinson proved that the origin-constrained problem is undecidable in the hyperbolic plane, see [14]. The general tiling problem for the hyperbolic plane remained pending a long time. In 2006, the present author proved that the tiling problem with an intermediate condition, so called **generalized origin-constrained problem** is undecidable, see [7,8]. Recently, the present author proved the general tiling problem to be undecidable in the hyperbolic plane, see [9,12]. At the same time, J. Kari established the same result, using a completely different approach, see [4].

In this paper, we prove that:

**Theorem 1** *The periodic domino problem is undecidable in the hyperbolic plane.*

The solution combines the construction given in [9,12] with an argument of [11] and a construction given in [6].

In the next section, section 2, we very sketchilly remind the solution to the tiling problem which we have given in [9,12,10]. In section 3, we prove the theorem.

## 2 The interwoven triangles

The solution of the domino problem which we now consider takes place in the tiling $\{7,3\}$ of the hyperbolic plane. It consists in delimiting infinitely many regions of infinitely many sizes in which the simulation of the same Turing machine is performed.

The construction takes place in the tiling $\{7,3\}$ of the hyperbolic plane, in which we define a particular tiling, the **mantilla**. In this tiling, we implement a construction which is based on what we call the **abstract brackets**, which is a construction on the line. We lift the intervals which are defined by the construction up to triangles in the Euclidean plane, with parallel legs whose heights lie on the same line. In this way, we can see the previous intervals as a projection of the triangles on the line of their heights. Then, we implement these triangles in the mantilla.

First, we sketchily describe the tiling $\{7,3\}$ of the hyperbolic plane and the mantilla.

### 2.1 The tiling $\{7,3\}$ of the hyperbolic plane

The tiling $\{7,3\}$ is obtained from the regular heptagon with an interior angle of $\frac{2\pi}{3}$ by reflection in its edges and, recursively, by reflection of the images in their edges. The existence of the tiling is a corollary of Poincaré's theorem on a sufficient conditions for tiling the hyperbolic plane by triangles. It is enough to consider the rectangluar triangle of the hyperbolic plane with the acute angles $\frac{\pi}{7}$ and $\frac{\pi}{3}$. Below, figure 2 illustrates the tiling $\{7,3\}$.

In [2], we introduced a way to exhibit a generating tree of the tiling which is basically the same as the generating tree of the pentagrid, the tiling $\{5,4\}$ of the hyperbolic plane. This tiling is constructed by a process, similar to the one used for constructing the tiling $\{7,3\}$, but it is used with the regular rectangluar pentagon. This tree is called the **standard Fibonacci tree**, simply **Fibonacci tree** in the sequel, see [5] for more details on this tree.

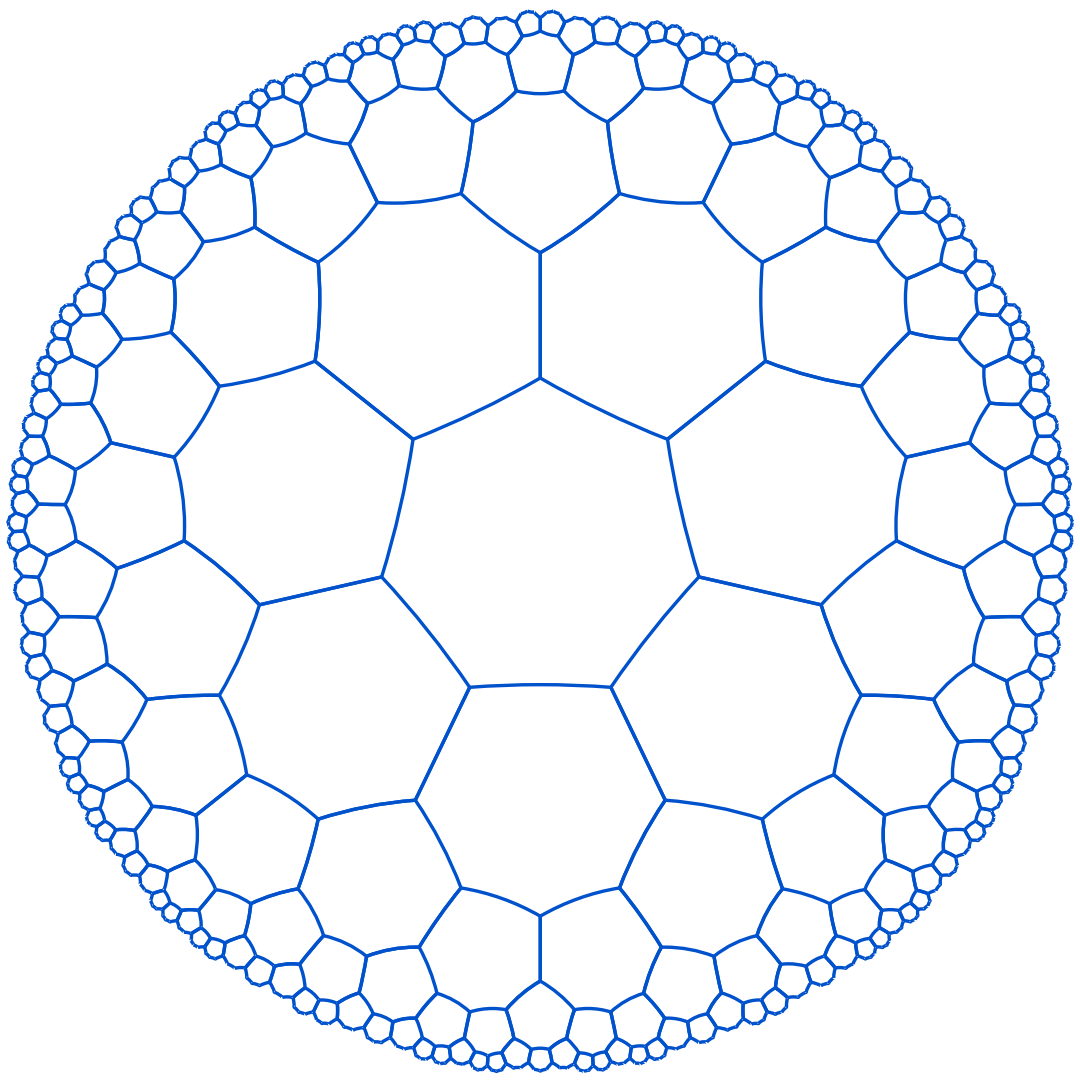

**Figure 1** *The tiling* $\{7,3\}$ *of the hyperbolic plane in the Poincaré's disc model.*

The way to exhibit the Fibonacci tree is based on the **mid-point lines**, which we introduced in [2]. As suggested by their name, these lines join the mid-points of two consecutive edges of a heptagon, see figure 2. It turns out that the angular sector determined by two rays obtained by two mid-point lines meeting at a mid-point $C$, joining the two mid-points of the two other edges which meet at $C$, exactly contains a set of tiles spanned by a Fibonacci tree. This structure will play an important rôle in what follows.

## 2.2 The mantilla

In the ternary heptagrid, a **ball** of **radius** $n$ around a tile $T_0$ is the set of tiles which are within distance $n$ from $T_0$ which we call the **centre** of the ball. The **distance** of a tile $T_0$ to another $T_1$ is the number of tiles constituting the shortest path of adjacent tiles between $T_0$ and $T_1$. We call **flower** a ball of radius 1.

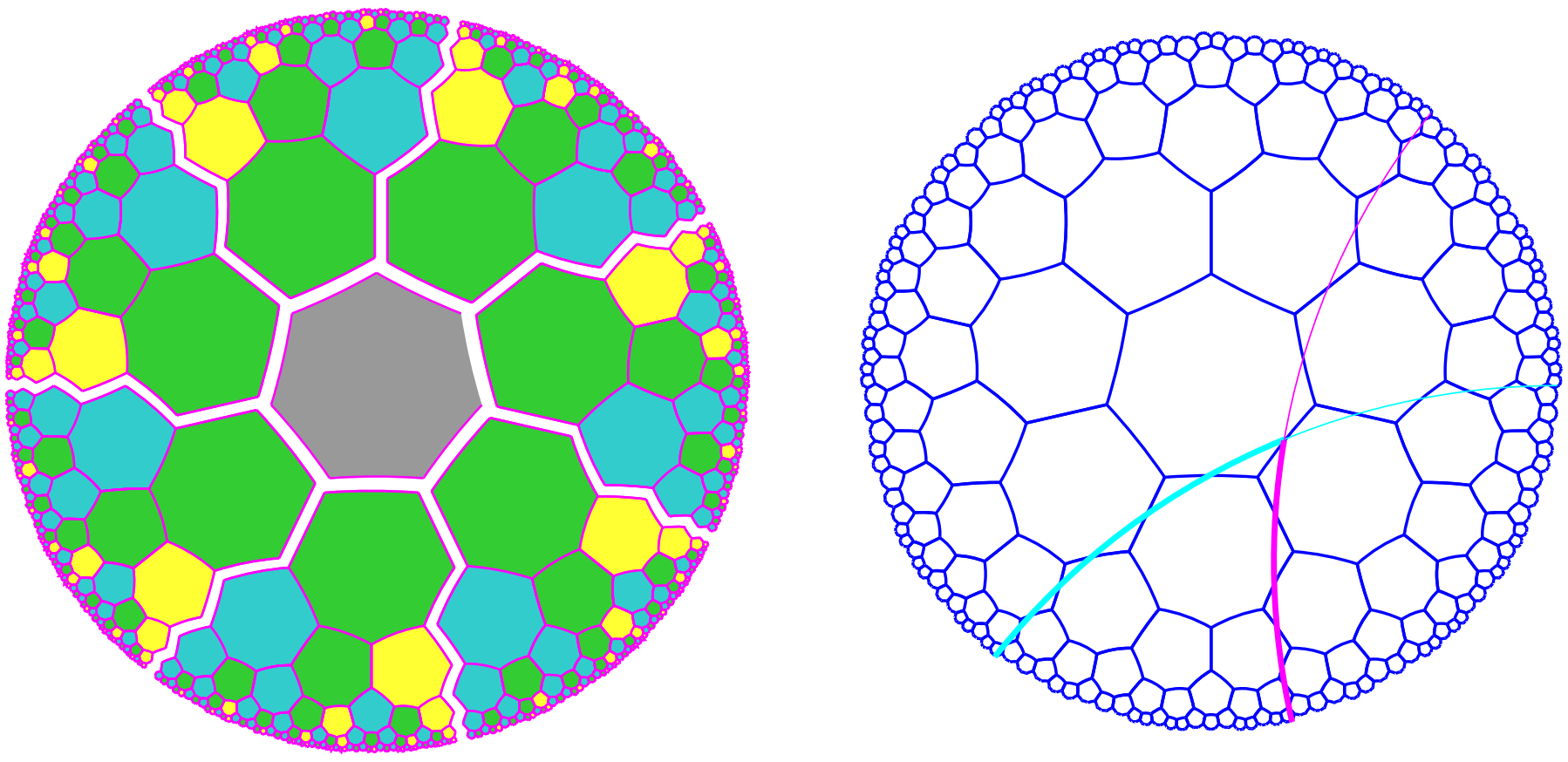

**Figure 2** *Left-hand side: the standard Fibonacci trees which span the tiling* $\{7,3\}$ *of the hyperbolic plane. Right-hand side: the mid-point lines.*

The mantilla consists in merging flowers in a particular way. Call **petals** of a flower the tiles which are around the central tile of the ball. The idea is that petals will be shared by flowers. In fact, each petal belongs to exactly three flowers. It is not very difficult to devise tiles which force such combinations. Figure 3 indicates the basic tiles for that purpose.

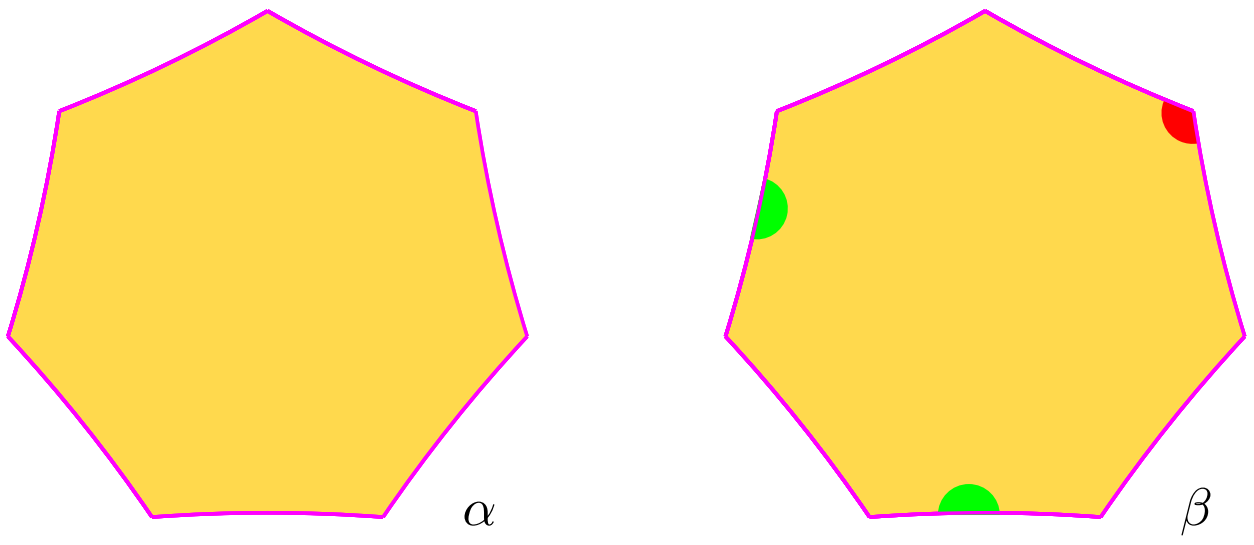

**Figure 3** *Left-hand side: the tile for the centres of the flowers. Right-hand side: the tile for the petals.*

Technically, we number the sides of the centres from 1 up to 7, turning clockwise or counter-clockwise around the tile. Of course, this induces corresponding marks on the edges of the petals which are in contact with a centre. The effect of such a numbering is to prevent tiles ($\alpha$) to alone tile the plane. With the numbering, we are forced to put tiles ($\beta$) around a tile ($\alpha$).

There are several types of flowers, depending on the position of the red vertices of the petals around the centre of the flower. Among the possible solutions, we select three kind of flowers, namely $F$-, $G$- and **8**-flowers which are below illustrated by figure 4.

In the same figure, we can see the **sectors** attached to each kind of flower. For $F$- and $G$-flowers, a sector is determined by the centre, the petals around the centre which are below the red vertices which are at distance 1 from the centre, and the two rays issued from these red vertices and which immediately cross an **8**-centre. For an **8**-flower, the sector is defined by the union of the four $F$-sectors which are just below the centre of the flower. Now, the sectors can be split in terms of $F$-, $G$- and **8**-sectors only. Figure 4 indicates these splittings.

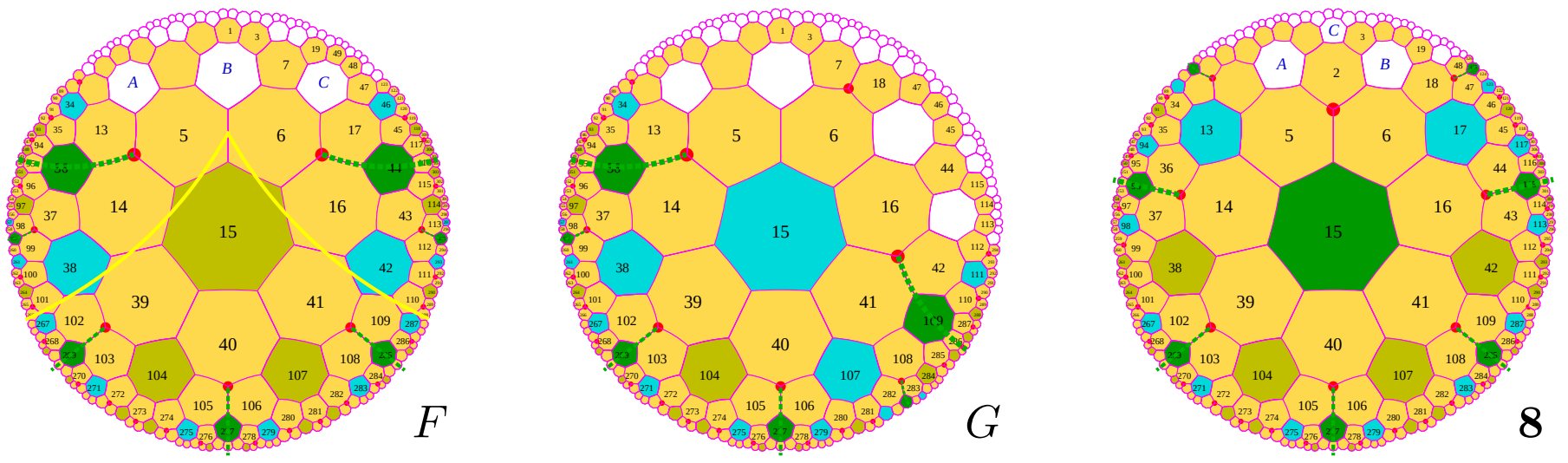

**Figure 4** *Splitting of the sectors defined by the flowers. From left to right: an $F$-sector, $G$-sector and* ***8****-sector.*

Now, we can define an important notion: say that an $F$-son of a $G$-flower is a **seed**. By definition, it is the root of a Fibonacci tree which is determined by the two rays issued from the mid-point of the edge shared by the two petals which contain the red vertices at distance 1 from the centre. The leftmost picture of figure 4 indicates the construction of such a tree which we call a **tree of the mantilla**. The set of tiles which are included in the angular sector determined by the rays is called the **area** of the tree, and the rays are called its **borders**. We shall also call border the set of tiles which are in contact with the rays. The trees of the mantilla have a very important property, see [7,8]:

1. *Consider two trees of the mantilla. Their borders never meet. Either their areas are disjoint or the area of one contains the area of the other.*

A second property allows us to define **horizontals** in this setting. This consists in considering that the seeds are black nodes in the sense of Fibonacci trees. As poved in [9,12], all other seeds inside such a tree are also black nodes and the **8**-centres contained in the tree are also black nodes. Number the edges of an $F$- or **8**-centre clockwise with the edges 1 and 7 between the two red vertices at distance 1 in an $F$-flower and sharing the single red-vertex in the case of an **8**-flower. Then, we connect the edges 2 and 6. When an $F$-flower is a white node, we connect the edges 1 and 6 or the edges 2 and 7. For $G$-flowers, we also number the edges from 1 to 7, but counter-clockwise. There are two kinds of $G$-flowers, depending on their position in the flower: one is on the left-hand side, the other on the right-hand side and they are called $G_r$- and $G_\ell$-flowers respectively. For $G_\ell$ flowers, we connect the edges 2 and 6, when it is a black node, the edges 1 and 6, when it is white. For a $G_r$-flower, we also connect the edges 2 and 6, when it is a black node, but the edges 2 and 7 when it is white. We call **isoclines** the maximal paths crossing exactly once consecutive connected edges and only them. The isoclines constitute the horizontals which will be used in the computing areas.

Now, we define the **verticals**. Note that an **8**-flower has a reflection axis which consists of the line which passes through the red vertex at distance 1 and through the mid-point of the centre which we call the **main** reflection axis of the **8**-flower. Call **verticals** the rays which start from the centre of an **8**-flower and go on along the main reflection axis of the **8**-flower and which do not pass through the red vertex of the flower. It is not difficult to see that a vertical belongs to the border of a sector. Now, verticals have an important property, see [9,12]:

2. *Consider a tree of the mantilla $T$. Let $v$ a vertical which starts from an* **8***-flower belonging to the area of $T$. Then, for any tree of the mantilla whose area $A$ is contained in that of $T$, $v$ does not meet $A$.*

### 2.3 The abstract brackets

By this name, we call the following process of construction of intervals on the line, illustrated by figure 5 which is performed by successive **generations**.

The generation 0 consists of points on a line which are regularly spaced. The points are labelled $R$, $M$, $B$, $M$, in this order, and the labelling is periodically repeated. An interval defined by an $R$ and the next $B$, on its right-hand side, is called **active** and an interval defined by a $B$ and the next $R$ on its right-hand side is called **silent**. The generation 0 is said to be **blue**.

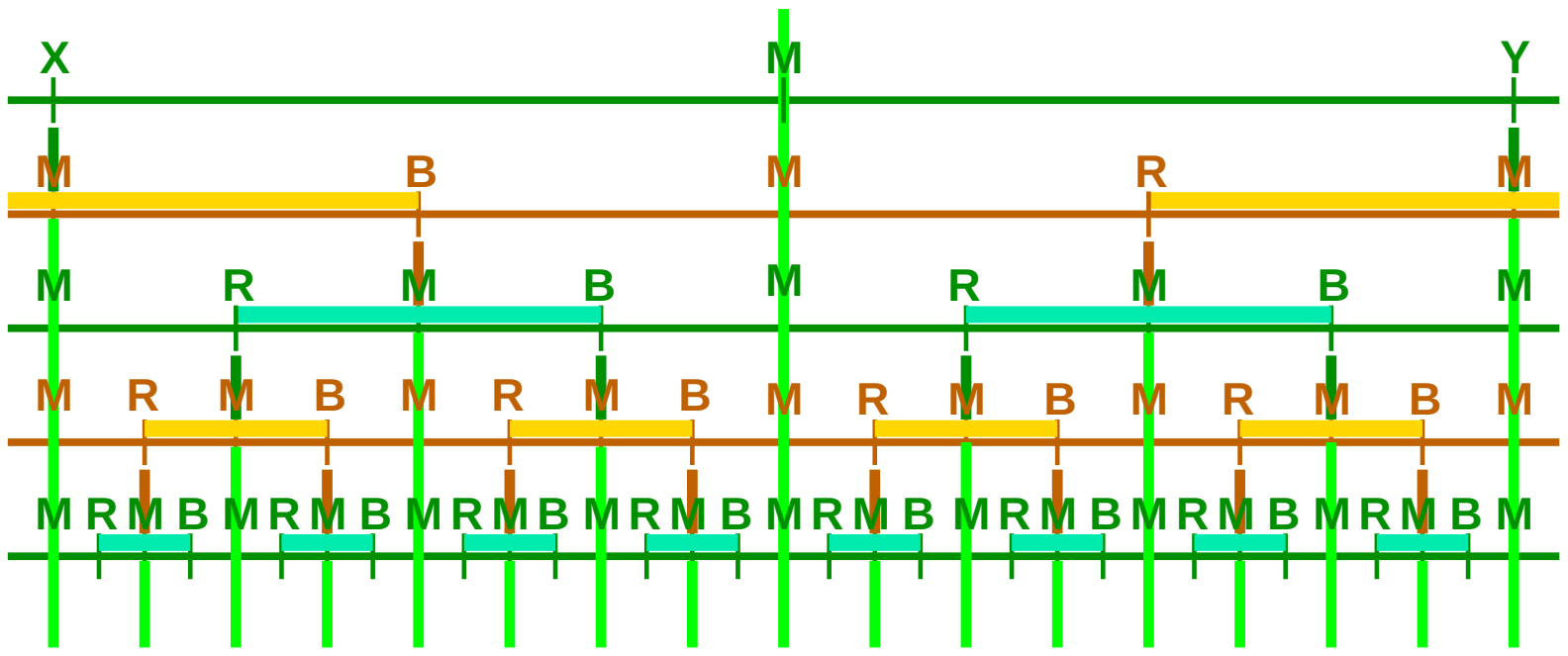

**Figure 5** *The silent and active intervals with respect to mid-point lines. The light green vertical signals send the mid-point of the concerned interval to the next generation. The colours are chosen to be easily replaced by red or blue inan opposite way. The ends $X$ and $Y$ indicate that the figure can be used to study both active and silent intervals.*

Blue and red are said **opposite**. Assume that the generation $n$ is defined. For the generation $n+1$, the points which we take into consideration are the points which are still labelled $M$ when the generation $n$ is completed. Then, we take at random an $M$ which is the mid-point of an active interval of the generation $n$, and we label it, either $R$ or $B$. Next, we define the active and silent intervals in the same way as for the generation 0. The active and silent intervals of the generation $n+1$ have a colour, opposite to that of the generation $n$.

When the process is achieved, we get an **infinite model**. The model has interesting properties, see [9,12,10].

In an interval of the generation $n$, consider that a letter of a generation $m$, $m \leq n$, which is inside an active interval is hidden for the generations $k$, $k \geq n+1$. Also, a letter has the colour of its generation. Now, we can prove that in the blue active intervals, we can see only one red letter, which is the mid-point of the interval. However, in a red active interval of the generation $2n+1$, we can see $2^{n+1}+1$ blue letters.

Cut an infinite model at some letter and remove all active intervals which contain this letter. What remains on the right-hand side of the letter is called a **semi-infinite model** which is also called a **cut** of an infinite model.

It can be proved that in a semi-infinite model, any letter $y$ is contained in at most finitely many active intervals, see [10].

### 2.4 The interwoven triangles in the Euclidean plane

As indicated in the introduction, we lift up the active intervals as **triangles** in the Euclidean plane. The triangles are isoceles and their heights are supported by the same line, called the **axis**, see figure 6.

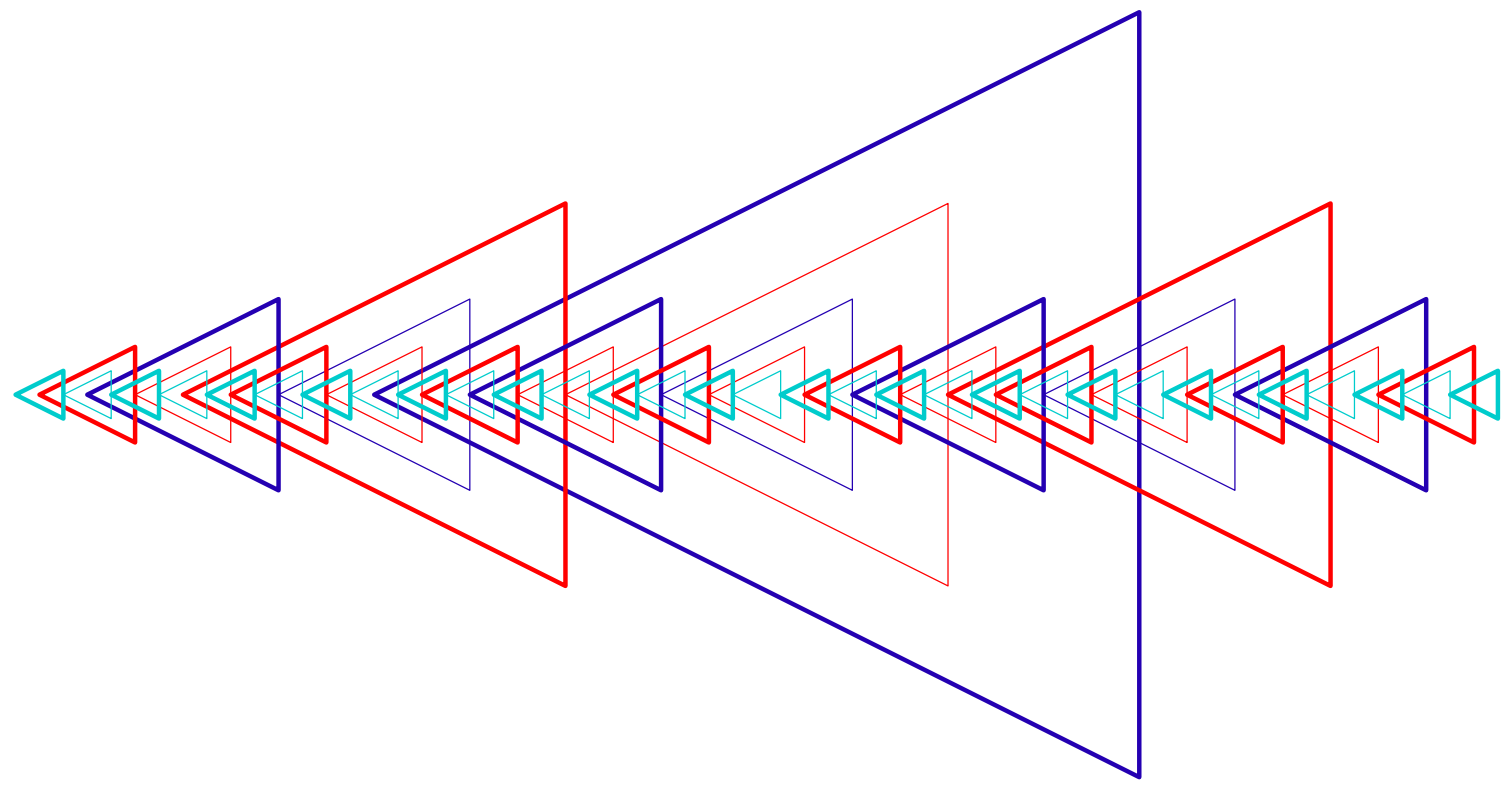

**Figure 6** *An illustration for the interwoven triangles.*

We also lift up silent intervals of the infinite model up to again isoceles triangles with their heights on the axis. To distinguish them from the others, we call them **phantoms**. We shall speak of **trilaterals** for properties shared by both triangles and phantoms.

We have very interesting properties for our purpose.

3. *Triangles of the same colour do not meet nor overlap: they are disjoint or embedded. Phantoms can be split into* **towers** *of embedded phantoms with the same mid-point and alternating colours. Trilaterals can meet by a basis cutting the half of leg which contains the vertex.*

From these properties, we prove in [10] that:

4. *The interwoven trilaterals can be obtained by a tiling of the Euclidean plane which can be forced by a set of* 190 *tiles.*

## 3 The tiling problem in the hyperbolic plane

We sketchily remember the implementation of the interwoven triangles in the hyperbolic plane.

To this purpose, note that the inclusion between areas of trees of the mantilla defines a partial order. We call **threads** the maximal sequences of totally ordered trees of the mantilla. Threads are indexed by $\mathbb{N}$ or by $\mathbb{Z}$. When it is indexed by $\mathbb{Z}$, the thread is called an **ultra-thread**. From [8], we know that two ultra-thread coincide, starting from a certain index. We also noticed in [8] that there can be realizations of the mantilla with ultra-threads as well as without them.

Now, in the hyperbolic plane, we implement different **cuts** of the **same** infinite model of the abstract brackets, taking a thread as the guideline for the realization of interwoven triangles.

The idea is that the triangles will be defined by the borders of trees of the mantilla as legs and by a piece of an isocline as the basis. Now, the isoclines will play the rôles of the letters in the abstract brackets.

In [10], we proved that if the isocline which passes through a seed $S$ is numbered 0, the closest seed inside the area of the tree defined by $S$ is on the isocline 5. Accordingly, we number the isoclines periodically from 0 up to 19. We select the isoclines 0, 5, 10 and 15 to play the rôles of the letters. These isocline are needed for the construction of the generations. Once the generation 0 is installed, the further generations are obtained by the algorithm suggested by the figures 5 and 6, see [9,12,10] for a precise description. Sketchily, the seeds grow legs until a green line is met. The green line is stopped by a triangle but not by a phantom. Afterwards, the leg go on growing until they meet the basis of their colour. For red triangles, we proceed, at the same time, to the detection of the **free rows** which are the isoclines corresponding to a free letter in a red active interval. For this process, the legs of the triangle diffuse a horizontal red signal outside the triangle and this, already from the generation 1. Accordingly, a row is free in a given red triangle $T$ if and only if there is no horizontal red signal on the considered isocline inside $T$.

Due to the presence of several trilaterals on the same interval of isoclines inside a given trilateral in the hyperbolic plane, it is needed to **synchronize** the processes which occur along different threads. In particular, when the threads merge, the processes may also merge, as they implement the same cut of the same infinite model.

To do this, we decide that triangles and phantoms of a given generation will have their basis and their roots on the same isoclines. Note that this is already the case of the generation 0. We simply extend this property to all the generations. To obtain this result, we decide that all bases of a given isocline merge into a unique basis signal which runs over the whole isocline. Now, the difference will be made by the presence of a horizontal upper signal of the colour of the trilateral, above the basis signal, on the same isocline. It may be realized as another channel on the same tiles, at a higher level than the channel used by the basis signal. Now, for coherence with the detection of the free rows in a red triangle, the vertices of the trilaterals also emit a horizontal signal of their colour but, this time, in a lower position: below the basis signal, again on the same isocline. The signals emitted by the legs of triangles at points which are neither the vertex nor the corner, are also of the colour of the triangle, but they are upper horizontal signals.

Moreover, the signals have a laterality given by the leg which they cross when they are of the same colour. Also, we allow signals of opposite lateralities to meet when they come from directions which are opposite to their lateralities. The other case of meeting is ruled out, except for the lower signals, where such a meeting is realized by the vertex itself. With these indications, a leg always know to which kind of basis it has to deal when it meets one of them.

Now, we select the red triangles and forget the others as well as the phantoms. The free rows and the verticals inside the triangle will be used to implement the space-time diagram of a Turing machine. As in Berger's and Robinson's proof, we consider the same Turing machine starting from an empty tape, whose simulation is performed in each computing area.

When the Turing machine does not stop, the computation is stopped by the basis of the concerned triangle, as its area is finite. As we have areas of infinitely many sizes, this allows to tile the plane. If the Turing machine halts, in one of the areas, the halting state will be called by the Turing machine. It is easy to associate to this state a tile which blocks the continuation of the tiling.

## 4 The periodic tiling problem

In [11], we proved that the finite tiling problem is undecidable. In fact, we shall use similar tiles to construct a periodic tiling of the hyperbolic plane when the simulated Turing machine halts. What is performed in [11] is that we have tiles which adapts to the halting tiles in order to encapsulate the computing area in a closed signal which runs along the legs and the basis of the concerned triangle.

Before going on, we shall remark an important property. Consider again an $F$-son of a $G$-centre. Then, replace the right-hand side ray by another one which delimits what we shall call a **black** tree. Such a tree is obtained by the same rules as for a standard Fibonacci tree, but its root is a black node. Technically, the root of a tree of the mantilla is also a black node, but everything happens as if we apply a white-node rule to the root and we change all white nodes which are on the right-hand side border into black node. In fact, a tree of the mantilla can be realized as a stack of black trees and standard Fibonacci trees which we shall call **white** trees to simplify the denotation. Now, with black trees we have the same properties as for the white ones: for any black trees of the mantilla, either their areas are disjoint or the area of one contains the area of the other. Accordingly, we can repeat the same construction process by replacing all trees of the mantilla by black trees. We refer the reader to [11] for a figure representing the tiles of such a border.

Accordingly, when the simulated Turing machine halts, we may construct two encapsulated areas $B$ and $W$ where the whole computation of the machine takes place. The area $W$ will be called white and the area $B$ will be called black. Moreover, from what we have noticed, both areas have the same height. Now, we shall make four super-prototiles with such computing areas. The border of an area has an interior side and an exterior one. It is not difficult to see that the difference can be noticed by the orientation of the tiles with respect to their father: number the edges of a tile counter-clockwise from 1 to 7, giving the number 1 to the edge which is shared by the father. Then the left-hand side border always go from the edge 1 to the edge 4 and the right-hand side border always go from the edge 1 to the edge 5. The difference between a black and a white tree is performed by the root. If the left-hand side border is also defined from the edge 1 to the edge 4 in the root, then the right-hand side border goes from the edge 2 to the edge 6 for a white tree and it goes from the edge 7 to the edge 4 for a black tree.

This distinction will allow us to look at the border as two-sided: one side is interior and the other is exterior. Now we shall consider that the left- and right-hand side borders of $B$ are black inside and white outside. Now, from $W$, we shall make three copies, $W_1$, $W_2$ and $W_3$. For $W_1$, $W_2$ and $W_3$, the inside of the borders is always white. For $W_1$, the outside of the left-hand and right-hand side borders is black. For $W_2$, the outside of the left-hand side border is black and that of the right-hand side border is white. For $W_3$, the outside of the left-hand side border is white and that of the righ-hand side border is black.

Now, we consider $B$, $W_1$, $W_2$ and $W_3$ as super-prototiles with which we shall construct a periodic tiling. In [6], we have considered sets of tiles which we called **quarters** and **bars**. A quarter of size $n$, denoted by $Q_n$, is the set of tiles spanned by a standard Fibonacci tree restricted to its first $n$ levels. A bar of size $n$, denoted by $R_n$, is defined in the same way as a quarter but with a black Fibonacci tree. In [6], we proved that $Q_{n+m}$ can be split into $Q_n$ and $f_{2n}$ copies of $Q_m$ and $f_{2n-1}$ copies of $R_m$. Note that $f_{2n}$ is the number of white nodes on the level $n$ of a standard Fibonacci tree and that $f_{2n-1}$ is the number of black nodes. In fact, the property comes from the fact that the trees rooted at two consecutive nodes of the same level of a Fibonacci trees are disjoint and that there is no node inbetween. Now, we can take $m = n$ and repeat the process as long as we wish. Using the property also proved in [6] that the hyperbolic plane can be viewed as the union of a growing sequence of quarters, we can construct the periodic tiling as follows.

*Initial step*:

Take a tile of the tiling $\{7, 3\}$ which will be called the **origin**. Take a copy of $W_2$, and place it in such a way that the root of the copy of $W_2$ coincides with the origin. Call this just defined region of the tiling $\mathcal{T}_0$. The root of this copy of $W_2$ is also called the top of $\mathcal{T}_0$ and its bottom border is called the bottom border of $\mathcal{T}_0$. Now, we define copies of $B$ and $W_i$, with $i \in \{1, 2, 3\}$ which we call $\mathcal{R}_0$ and $\mathcal{Q}_0^i$ respectively. We define the roots and the bottom borders of these regions as we did for $\mathcal{T}_0$.

Now, in what follows, a particular rôle will be played by the node of the last level of $W_2$ whose coordinate is a term of the Fibonacci sequence. Call it the **junction point**. Now, on the tiling $\{7,3\}$, draw the line which passes throught the mid point of the origin and through the junction point of $\mathcal{T}_0$. Call it the **axis**. This line is simply an auxiliary tool in our construction.

*Induction step*:
Assume that $\mathcal{T}_n$ is constructed, as well as regions $\mathcal{R}_n$ and $\mathcal{Q}_n^i$, which have the following particularity: the left- and right-hand side borders of $\mathcal{Q}_n^i$ also have an outside and an inside parts and the outside and inside part of the left- and right-hand side borders of $\mathcal{Q}_n^i$ have the same colour as the correpsonding elements of $W_i$, for each $i \in \{1,2,3\}$.

Then, take a copy of $W_2$ and put it above $\mathcal{T}_n$ in such a way that the top of $\mathcal{T}_n$ is the middle son of the junction point of $W_2$. Now, on the bottom border of $W_2$, proceed as follows, starting from the leftmost node: if we have a black node, place a copy of $\mathcal{R}_n$, then a copy of $\mathcal{Q}_n^1$; if we have a white node, place copies of $\mathcal{R}_n$, $\mathcal{Q}_n^2$ and $\mathcal{Q}_n^3$, in this order. This is the first step: we get a set of tiles which is alike a copy of $Q_m$ for an appropriate $m$. Now, consider the bottom border of this region, and proceed as follows, starting from the leftmost node: if we have a black node, place a copy of $B$, then a copy of $W_1$; if we have a white node, place copies of $B$, $W_2$ and $W_3$, in this order. Now, we get a new region $\mathcal{T}_{n+1}$ which strictly contains $\mathcal{T}_n$: the tiles which belong to the border of $\mathcal{T}_n$ do not meet the tiles which belong to the border of $\mathcal{T}_{n+1}$. Now, we construct $\mathcal{R}_{n+1}$ and $\mathcal{Q}_{n+1}^i$ in a similar way, starting from, respectively $\mathcal{R}_n$ and $\mathcal{Q}_n^i$ and completing them by two rows of copies of $B$ and $W_i$'s which are placed as above indicated.

Now, it is plain that $\underset{n\in N}{\cup} \mathcal{T}_n$ is the hyperbolic plane. Also, from the construction, it is plain that this tiling is invariant under the shift along the axis which transforms the origin in the middle son of the junction point of $\mathcal{T}_0$.

To complete the proof, assume that the simulated Turing machine does not halt.

If there is a solution, the tiles for the border of the super-tiles always admit near them a seed $S$ which is of the generation 0: hence, it does not bear the root of a copy of a super-tile. Next, we know that the computation goes on endlessely, thanks to the construction of [9,12,10]: as the halting state cannot be met, the occurrence of tiles of the border of super-tiles cannot be triggered inside the tree of the mantilla rooted at $S$. If the tiling would be periodic, then from the invariance under a shift, we would easily get that as $S$ does not contain tiles of the border of a super-tile, this is also the case for the whole plane. Now, this means that we have a solution of the tiling constructed in [9,12,10] and, as we know, such a solution cannot be periodic. Indeed, a shift should keep the isoclines globally invariant and then, there is no shift which would match the triangles of a certain generation and those which are of a higher generation. But this contradicts the assumption of a periodic solution. Accordingly, if the Turing machine does not halt, there is no periodic solution. The just produced argument indicates that in this case there are solutions, but they are not periodic.

This completes the proof of theorem 1. ■

## 5 Conclusion

I hope that this paper shows the interest of the construction given in [9,12,10]. As indicated in [11], there is still much work to do in this domain.

## References

1. Berger R., The undecidability of the domino problem, *Memoirs of the American Mathematical Society*, **66**, (1966), 1-72.
2. Chelghoum K., Margenstern M., Martin B., Pecci I., Cellular automata in the hyperbolic plane: proposal for a new environment, *Lecture Notes in Computer Sciences*, (2004), **3305**, 678-687, (proceedings of ACRI'2004, Amsterdam, October, 25-27, 2004).
3. Gurevich Yu., Koriakov I., A remark on Berger's paper on the domino problem, *Siberian Mathematical Journal*, **13**, (1972), 459–463.
4. Kari J., A new undecidability proof of the tiling problem: The tiling problem is undecidable both in the Euclidean and in the hyperbolic plane, *AMS meeting a NC, Special Session on Computational and Combinatorial Aspects of Tilings and Substitutions*, March, 3-4, (2007).
5. M. Margenstern, New Tools for Cellular Automata of the Hyperbolic Plane, *Journal of Universal Computer Science* **6**N°12, 1226–1252, (2000)
6. M. Margenstern, Cellular Automata in Hyperbolic Spaces, Volume 1, Theory, *OCP*, Philadelphia, to appear (2007).
7. Margenstern M., About the domino problem in the hyperbolic plane from an algorithmic point of view, *arxiv:cs.CG/*060393, (2006), March, 11pp.
8. Margenstern M., About the domino problem in the hyperbolic plane from an algorithmic point of view, *Publications du LITA, N°*2006-101*, Université de Metz*, (2006), 110pp.
9. Margenstern M., About the domino problem in the hyperbolic plane, a new solution, *arxiv:cs.CG/*070196*v*1*, arxiv:cs.CG/*070196*v*2, (2007), January, 60pp.
10. Margenstern M., About the domino problem in the hyperbolic plane, a new solution, *Publications du LITA, N°*2007-102*, Université de Metz*, (2007), 107pp.
11. Margenstern M., The finite tiling problem is undecidable in the hyperbolic plane, *arxiv:cs.CG/*0703147, (2007), March, 8p.
12. Margenstern M., A tiling of the hyperbolic plane: the mantilla, with an application to the tiling problem, *AMS meeting at Davidson, NC, Special Session on Computational and Combinatorial Aspects of Tilings and Substitutions*, March, 3-4, (2007).
13. Robinson R.M. Undecidability and nonperiodicity for tilings of the plane, *Inventiones Mathematicae*, **12**, (1971), 177-209.
14. Robinson R.M. Undecidable tiling problems in the hyperbolic plane. *Inventiones Mathematicae*, **44**, (1978), 259-264.
15. Wang H. Proving theorems by pattern recognition, Bell System Tech. J. vol. **40** (1961), 1–41.